# Meta-Diastereomers Hierarchical Multiscale Chiral Interactions Between Biomolecules and Nanoscale Enantiomers.

Dominic J.P. Koyroytsaltis-McQuire[1], Rahul Kumar[1], Shailendra K Chaubey[1], Tamas Javorfi[2], Giuliano Siligardi[2], Affar Karimullah[1] Adrian J Lapthorn[1], Nikolaj Gadegaard[3], Malcolm Kadodwala[1*]

[1] School of Chemistry, University of Glasgow, Glasgow, G12 8QQ, UK

[2] Diamond Light Source Ltd., Harwell Science and Innovation Campus, Didcot, OX11 0DE, UK

[3]School of Engineering, Rankine Building, University of Glasgow, Glasgow G12 8LT, UK

* Corresponding Authors

E-mail: Malcolm.kadodwala@glasgow.ac.uk

# Abstract

We introduce meta-diastereomers, hybrid systems where molecular chirality (≤10 nm) and nanoscale chirality (>100 nm) combine to create a new hierarchical chiral state with emergent optical properties. By coupling chiral biomolecules electrostatically with enantiomeric silicon nanostructures, we demonstrate how distinct chiral systems can interact to produce optical responses—linear and circular dichroism—that reflect the combined chirality of both components. This unique interplay of chirality across scales represents a conceptual advance in chiral nanophotonics. Importantly, we leverage the meta-diastereomers to probe a model antibody-antigen interaction, analogous to how molecular diastereomers are used in chemistry to investigate chiral structures. This work establishes a platform for label-free, highly specific detection of biomolecular interactions and opens new avenues for exploring hierarchical chirality in optics, biosensing, and quantum technologies.

# Introduction.

Chirality—the property of an object being non-superimposable on its mirror image—is a fundamental phenomenon that underpins processes in chemistry[1], biology[2], and optics[3]. At the molecular scale, chirality determines the biological activity of drugs, the structure of proteins, and the stereochemical behaviour of materials. At the nanoscale, chiral inorganic materials can manipulate the spin of electrons and the polarization of light, enabling breakthroughs in photonics, quantum technologies, and sensing. However, while both molecular and nanoscale chirality have been extensively studied, these systems have traditionally been treated in isolation, with little attention given to their interplay.

A critical and unresolved question is: How can chirality at two distinct length scales—molecular (≤10 nm) and nanoscale (>100 nm)—interact and combine? Addressing this question could provide not only fundamental insights into hierarchical chirality but also open pathways for new technologies that exploit combined chiroptical responses. Previous studies have primarily investigated molecular or nanoscale chirality in isolation[4-21] or their weak interactions via near-field effects[22-24]. However, these approaches have not explored the mechanisms by which molecular chirality and nanoscale structural chirality can combine to form a new chiral entity with emergent, combined chiroptical properties

Here, we introduce "meta-diastereomers"—hybrid systems that combine the chirality of biomolecules with the intrinsic chirality of nanoscale silicon nanostructures through electrostatic coupling. Meta-diastereomers represent a new hierarchical chiral state, where molecular chirality (≤10 nm) and nanoscale chirality (>100 nm) interact synergistically to produce emergent chiroptical responses, including distinct linear and circular dichroism. These optical signatures provide direct evidence of the combined contributions of molecular and nanoscale chirality, and the creation of an entity which as diastereomeric properties.

We demonstrate the functional significance of meta-diastereomers by applying them as optical chiral probes to detect a model antibody-antigen interaction. This functional demonstration draws an analogy to the established role of molecular diastereomers in probing stereochemical environments and highlights their potential for label-free, highly specific biosensing of biomolecular interactions.

Our work establishes a conceptual and experimental framework for combining chirality at multiple length scales through electrostatic interactions, introducing meta-diastereomers as a new class of hierarchical chiral systems. By demonstrating their unique chiroptical properties and biosensing capabilities, this approach paves the way for the design of chiral photonic devices and advances the understanding of multiscale chirality in functional nanotechnology.

# Results & Discussion

## Meta-Diastereomer Formation through Electrostatic Coupling

In this study, meta-diastereomers are created from chiral elements, silicon asymmetric double split-ring resonators with "S"-shaped geometries. The structural elements were intended to have a nominal thickness of approximately 160 nm; however, due to variations in the fabrication process, the actual thicknesses varied from this value by ±10 nm between samples. The lateral dimensions and AFM images of the structures are shown in Figure 1a and 1b. These structures exhibit chirality due to the breaking of horizontal mirror symmetry caused by the glass substrate. As members of the $C_1$ point group, the S-shaped structures display uneven spacing between the top and bottom arms and the middle arm (Figure 1a).

Four distinct metasurfaces were developed from these chiral elements, organized into square arrays with a periodicity of 850 nm (Figure 1b) with each array consisting of 600 $\times$600 nanostructures. These arrays comprised both left-handed (LH) and right-handed (RH) enantiomorphs, along with two racemic configurations (RA and RS). The RS configuration features alternating LH and RH structures, which are clearly visible in the AFM image (Figure 1b). In contrast, the RA configuration contains alternating domains composed of 19x19 sub-arrays, each consisting exclusively of either LH or RH structures. The variations in thickness across different samples resulted in slight differences in the optical properties from samples fabricated at different times. Additionally, the absence of four-fold rotational symmetry in all silicon metasurfaces leads to inherent birefringence. The optical properties of these metasurfaces have been previously studied in detail[25]

The metasurfaces were functionalised with streptavidin conjugated to quantum dots (Qdot™ Thermo Fisher), denoted as strept, through a well-established multistep process[26, 27]. Luminescence emitted by the QDs was utilized to confirm streptavidin immobilization. Initially, the sample was immersed in PBS buffer for at least 5 minutes, followed by the application of the first species, poly-L-lysine (PLL). PLL, a standard pre-coating reagent, facilitated the adhesion of biomolecules and cells, forming a robust monolayer that allowed extensive rinsing without notable coverage loss[28, 29]. The samples were kept in PLL for 15 minutes, enabling the positively charged PLL to adhere to the silicon surface. Subsequently, glutaraldehyde was introduced and left for 45 minutes to react with the amine groups of the PLL layer. Strept was then added for 30 minutes, undergoing a cross-linking to the PLL layer[30, 31]. To block unbound glutaraldehyde sites and reduce nonspecific binding, 40 mmol of ethanolamine was applied as a capping layer. After 20 minutes, the solution was replaced with PBS, Figure 1c. When required these functionalised metasurfaces were incubated with PBS buffered solutions of polyclonal anti-streptavidin (anti-strep) for two hours. After incubation the metasurfaces were rinsed with PBS and spectra were collected with the substrates immersed in buffer.

## Optical Evidence of Hierarchical Chirality and its Application to Probing Antibody-Antigen Interactions.

**Unfunctionalised metasurfaces:** The reflectance spectra of the unfunctionalised metasurfaces are consistent with earlier findings (Figure 1a)[25]. The spectra exhibit a prominent doublet resonance centred around 655 nm, along with a less pronounced resonance near 700 nm. This distinctive spectral pattern was consistently observed across all metasurface designs. Previous numerical simulations incorporating multipolar analysis identified these resonances as predominantly arising from magnetic dipole contributions[25], which is a recognised feature of dielectric nanostructures with engineered geometries.

To expand the understanding of these metasurfaces the dependency of the optical response on dielectric properties (refractive index (RI)) of the surrounding environment were investigated, Figure 2a. By comparing the spectra obtained from metasurfaces immersed in phosphate-buffered saline (PBS) (RI = 1.339) with those in racemic 2-butanol (RI = 1.397), a small red shift of approximately 3 nm was observed. This corresponds to a metasurface sensitivity of approximately 48 nm per refractive index unit (RIU). This low sensitivity, particularly in comparison to plasmonic surface plasmon resonances (SPR), which can attain sensitivities in the thousands[32, 33], clearly demonstrating the relative insensitivity of optical properties to the RI of surrounding dielectric environment.

**Functionalised metasurfaces**: the deposition of biomolecules markedly perturbs the reflectance spectra of all metasurfaces (Figure 2b), with a consistent response across all four types. Unexpectedly, biomolecule adsorption induces a blue shift in resonances, contrary to the anticipated red shift associated with an increased local refractive index. Streptavidin deposition results in both a blue shift and a decrease in resonance intensity. Subsequent binding of anti-streptavidin to the streptavidin layer causes further, albeit smaller, blue shifts and additional reductions in resonance intensity.

A useful comparison for the current study is the work on the influence of adsorbed charged species on Mie scattering from spherical achiral dielectric particles[34-38]. This work also revealed a blue shifting in scattering behaviour. Thus, the observed blue shift in the optical resonance can be considered a fingerprint of an electrostatic coupling between the dielectric chiral nanostructures and charged biomolecules

To exclude the possibility of morphological changes in the nanostructure as the origin of these spectral alterations, AFM imaging was conducted post-biomolecule deposition. These images revealed no appreciable changes in the height or shape of the S-shaped structures, eliminating significant structural modifications or damage as potential causes of the observed spectral shifts (see Supplementary Information S1).

Stokes' polarimetry is commonly used for rapid analysis of light polarization in metamaterial studies, however, it lacks the ability to fully describe light-matter interactions and differentiate between optically active (chiral) and birefringent effects. In contrast, Mueller Matrix Polarimetry (MMP) provides a more detailed analysis using a 4×4 Mueller matrix, offering a comprehensive understanding of how light polarization is altered during interaction with matter. This method for non-depolarizing materials enables the separation of birefringent and optically active (chiral) responses[39]. While enantiomer pairs exhibit identical linear dichroic responses, they can be distinguished by their equal and opposite optically active responses. Conversely, diastereomer pairs demonstrate non-identical, yet oppositely signed, optically active responses. Crucially, for the current study, linear dichroism can be employed to differentiate between these diastereomers.

The spectra of the 16 Mueller elements, $M_{ij}$ where i and j range from 0 to 3, for the unfunctionalized, streptavidin, and anti-streptavidin cases, are provided in the supplementary information (S2). The symmetry relations between the spectral responses of the $M_{ij}$ elements are governed by the symmetry properties of the medium interacting with the light. In the case of all LH and RH the $M_{03}$ and $M_{30}$ elements display similar line shapes but opposite sign. This is consistent with low symmetry materials which lack fourfold rotational axes[39, 40], and would be expected for the S structures. Such materials cause depolarisation of circularly polarised light which is asymmetric and non-reciprocal. Consequently, CD response of these low symmetry materials have both a reciprocal chiral and a non-reciprocal nonchiral contribution. This later contribution does not change sign on switching between enantiomers[39, 40].

The following spectra can be derived from the $M_{ij}$ spectra: linear dichroism (LD, LD'), circular dichroism (CD), circular birefringence (CB), and linear birefringence (LB, LB') spectra. LD and LD' measurements revealed extinction differences for light polarized at 0°/90° and ±45°, respectively. CD and CB spectra were sensitive to the optical activity of chiral media, while LD and LB measurements captured birefringent response. For brevity we have focused on the linear and circular dichroic data.

In general, although LD' and CD spectra have superficial similarities, displaying equal and opposite spectra for enantiomorphic behaviour, LD' measurements are non-reciprocal – the spectra switch sign when either the direction of propagation is flipped or the sample is rotated by 90°. This indicates that LD' is not a chirally sensitive measurement, unlike CD, which must be reciprocal.

Shown in Figure 3a-c are LD, LD', and CD spectra acquired from unfunctionalized, strep, and strept-anti-strept functionalized LH and RH enantiomorphic metasurfaces. The spectral patterns align with the reflectance data, showing a progressive blue shift in the resonances following both streptavidin deposition and subsequent anti-streptavidin binding. As with reflectance, the dominant features in LD, LD', and CD spectra correspond to the magnetic dipole resonance. Biomolecule adsorption induces asymmetric changes in the magnitudes of the magnetic dipole resonances in LD, LD', and CD spectra for enantiomorphic substrates. Asymmetry parameters ($\alpha^{CD,LD,LD'}_{Buffer,Strept,Anti-Strept}$) for unfunctionalized, Strept and anti- layers between LH and RH handed structures derived from the magnitude of the dominant peak in the spectra (labelled R in figure 3a-c) which is associated with the magnetic dipole response,

$$\alpha^{CD,LD,LD'}_{Buffer,Strept,Anti} = \frac{\left|I^{CD,LD,LD'}_{Buffer,Strept,Anti-Strept}(LH)\right|}{\left|I^{CD,LD,LD'}_{Buffer,Strept,Anti-Strept}(RH)\right|}$$

Where $I^{CD,LD,LD'}_{Buffer,Strept,Anti-Strept}(LH/RH)$ are the magnitude of the signal at peak R. Values of presented in Table 1. The asymmetry trends are consistent across all three measurements: streptavidin deposition induces significant asymmetries, which decrease upon anti-streptavidin binding. Most significantly the LD spectra of the enantiomers go from being identical to showing significant difference after deposition of streptavidin. This is a signature of diastereomer formations, as pairs of enantiomers give the same response. The diastereomer effect becomes much weaker when anti-strept is bound to the strept.

It should be noted that in the ideal case the asymmetry parameters for the buffer should be 1. One may expect the asymmetry parameters for the buffer to diverge from unity due to differing levels of structural heterogeneity. However, $\alpha^{CD}_{Buffer}$ shows a greater deviation from unity than the equivalent values for LD and LD′. This is attributed to the non-chirally sensitive contribution to the CD due to the depolarising properties of the S structures.

The dichroic spectra from the racemic structures, Figure 4 and supplementary information S3, display behaviour consistent with that expected based on the response of the enantiomorphs. It should be noted that the presence of the binding of strept, and subsequent binding of anti-strept, causes a change in the dichroic responses.

To summarise, it is proposed that the induction of asymmetry in the LD responses of LH and RH metasurfaces after strept binding is a signature of diastereomer formation. The subsequent binding of anti-strept reduces the diastereomer effect.

To validate experimental observations and gain insights into the mechanism responsible for the formation of the meta-diastereomer, we conducted numerical simulations using the COMSOL Multiphysics platform. Our simulations aimed to rationalize two primary experimental observations: 1) a blue shift, and 2) the enantiomorphic dependency of both optical activity and birefringent responses after biomolecule adsorption. This section specifically focuses on the modelling of individual enantiomorphic metasurfaces. For details regarding the modelling of racemic metasurfaces, please refer to the supplementary information section S5.

The numerical modelling employed an idealized S structure, deliberately neglecting structural, morphological, and compositional heterogeneities, such as dopants introduced by contamination that could modify charge carrier densities or the oxidation of the silicon surface.

**Modelling of Rac-2-butanol**: To confirm the robustness of our modelling procedure, we simulated the reflectance spectra of the enantiomorphic structure immersed in water and rac-2-butanol, as depicted in Figure 5a. The simulated data exhibited very good agreement with experimental results, accurately reproducing both the spectral line shape and indicating a slight ∼3 nm redshift in the simulated reflectance spectra between buffer and 2-butanol. This concordance with experimental findings affirms the accuracy and reliability of the modelling procedure.

**Modelling of Charged Proteins**: The experimental blue shift observed could not be explained within the conventional refractive index framework, where the addition of a protein layer (dielectric layer) and consequent local increase in refractive index results in a red shift. To investigate the hypothesis that the observed blue shifts is associated with the electrostatic interaction generated by surface dipoles associated with charged proteins, figure 6a-b, we introduced a dipole layer around the S-shaped structures. Adapting the model used for simulating buffer and rac-2-butanol spectra, we included 20 nm chiral dielectric domains around the outer edges of the nanostructure, Figure 6c. This dielectric layer, assigned a refractive index of 1.4, exhibited chirality reflected in the Pasteur coefficient (ξ) of $1.7 \times 10^{-4}$, consistent with values used in previous studies to model protein layers[41-44].

To account for surface charge effects from deposited molecules, we introduced an external polarization to the dielectric layer, figure 6c. This was approximated as a dipole layer extending from negatively charged streptavidin to positively charged PLL, which is directly bound to the surface. The negative charge on streptavidin is attributed to the pH of the buffer being above the isoelectric point pI) of the protein (pI ~ 5 – 6)[45, 46]. An estimate for the realistic polarization magnitude, based on streptavidin dimensions and a previously reported dipole moment, yielded a polarization of $8 \times 10^{-4}$ $Cm^{-2}$ [47]. However, for qualitative agreement with the experiment, a value of $1 \times 10^{-8}$ $Cm^{-2}$ was utilized in simulations. The necessity to use a lower value for polarization to obtain qualitative agreement with the experiment can be rationalized in part by considering the Debye shielding effects arising from the presence of ionic species within the buffer solution.

To evaluate the modelling procedure's robustness, individual simulations were conducted in a stepwise manner, systematically introducing layer properties. The sequence began with the refractive index, followed by ξ ($\pm 1.7 \times 10^{-4}$), and concluded with the polarization. Switching the sign of the ξ is equivalent to switching enantiomer, due to symmetry it is also equivalent to having the same enantiomer but switching the handedness of the metasurface. Simulated reflectance spectra from each step are depicted in Figure 5b-c. The introduction the of an achiral unpolarised dielectric layer refractive index 1.4, results in the expected red shift ~ 3 nm. Making this dielectric layer by assigning it a ξ value has a negligible effect, with slight asymmetry between the positive and negative values of ξ ($\pm 1.7 \times 10^{-4}$). However, the inclusion of a static polarisation induces a “split-shoulder” to the blue of the magnetic dipole resonance at ~665 nm. Equally importantly, switching the sign of ξ results in a significant asymmetry in the intensity of the resonances associated with the magnetic dipole.

We acknowledge a limitation in simulation, arising from the use of a simplistic model for statically polarized layers that introduces some physically unrealistic aspects. Specifically, at both sides of the structure's edges, the orientation of polarizations is perpendicular, leading to a discontinuity in the gradient of the static field. This could be origin of quantitative differences between simulated and experimental spectra.

To offer further insights, field maps illustrating the electric field intensity |E| and the chiral asymmetry, parameterized by the optical chirality factor C[48, 49], in the electromagnetic (EM) environment are presented in Figure 7a-b. In instances where a dielectric layer solely possesses static polarizations, no asymmetries are observed in either |E| or C between the two enantiomorphs. However, in the case of a chiral dielectric with static polarizations, asymmetries emerge in both |E| and C within the internal EM environment of the silicon structure.

Considering that linearly polarized light induces asymmetries in |E| in the internal EM environment, which, in turn, triggers asymmetries in scattering and absorption, this observation supports the noted asymmetry in the birefringent response. Concurrently, the asymmetries in C align with the observed asymmetries in the optical active response.

The final consideration revolves around whether the proposed model can elucidate the decrease in the magnitude of asymmetries when anti-streptavidin binds to a functionalized surface. The binding of anti-streptavidin will have a dual impact, increasing the thickness and as will be argued below, simultaneously reducing the overall polarization of the dielectric protein layer. Both effects collectively influence the modification of the optical response of the metasurfaces upon antibody binding.

The anti-streptavidin employed in this study is a polyclonal antibody, implying its interaction with a variety of epitopes on the streptavidin layer. Consequently, the bound anti-streptavidin will adopt a relatively broad range of adsorption geometries, causing a decrease in the anisotropy of the alignment of the dipole moment within the antigen-antibody complex. This, in turn, results in a reduction in the magnitude of the net dipole moment and the polarization of the layer, as depicted in Figure 8. The diminished polarization, consequently, lowers the level of asymmetry in the birefringent and optically active responses. It is noteworthy that the binding of anti-streptavidin still induces a blue shift, albeit smaller than that induced by the functionalization with streptavidin. Thus, the electrostatic model can partially rationalize the observed effects of antibody binding. However, the increased structural complexity of an antibody-antigen complex undoubtedly goes beyond the simplistic model discussed here.

# Conclusion

In this work, we introduced meta-diastereomers, a new class of hierarchical chiral systems formed through the electrostatic coupling of molecularly chiral biomolecules and nanoscale enantiomeric silicon structures. By combining chirality at two distinct length scales—molecular (≤10 nm) and nanoscale (>100 nm)—we demonstrated the emergence of synergistic optical responses, including distinct linear dichroism (LD) and circular dichroism (CD) signals. These responses provide direct evidence for the combined contributions of molecular and nanoscale chirality, establishing meta-diastereomers as a robust framework for hierarchical chiral systems.

The functional significance of this approach was highlighted through the successful demonstration of meta-diastereomers as optical chiral probes to detect subtle changes in molecular chirality during an antibody-antigen interaction. This application underscores their potential for label-free, highly specific biosensing of biophysical interactions, bridging chiral nanotechnology with real-world diagnostics.

By addressing the long-standing challenge of combining chirality across scales, our work challenges conventional paradigms that treat molecular and nanoscale chirality as independent phenomena. It opens pathways for the design of chiral photonic devices, materials with tailored chiroptical properties, and advanced nanotechnologies for biosensing and quantum applications.

# Methods & Experimental

### Sample fabrication

Sample fabrication was performed at the James Watt Nanofabrication Centre (JWNC). The s-structures were created using electron beam lithography. Quartz glass slides were first cleaned by ultrasonic agitation in acetone, methanol, and isopropyl alcohol (AMI) for 5 minutes each. Following cleaning, the slides were dried under a nitrogen flow and subjected to an oxygen plasma treatment for 5 minutes at 100 W. Amorphous silicon was then deposited onto the substrates using Plasma Enhanced Chemical Vapor Deposition (PECVD) with an SPTS Delta tool. The samples were subsequently cleaned again, but with a 1-minute low-power (60 W) plasma treatment.

Next, a bilayer of PMMA resist was spun onto the substrates at 4000 rpm for 1 minute and baked at 180 °C for 5 minutes between each spin. A 10 nm aluminium conducting layer was deposited using a PLASSYS MEB 550s evaporator. Patterns were designed using L-Edit CAD software and written with a Raith EBPG 5200 electron beam lithography system operating at 100 kV. The aluminium was removed using CD-26, and the resist was developed in a 3:1 mixture of MIBK: IPA at 23.2 °C for 1 minute, then rinsed in IPA for 5 seconds and deionised water before being dried under nitrogen flow. A 50 nm nichrome etch mask was then deposited on the sample. This was followed by a lift-off procedure in acetone at 50 °C overnight, with the sample agitated to remove any residual resist and excess metal. The silicon was etched using an STS tool with a custom recipe of C4F8 / SF6 (90/30 sccm), 600 W, 9.8 mTorr, 20 °C for 3 minutes. The NiCr etch mask was removed by immersion in chromium etchant and 60% nitric acid. The final steps involved an AMI and low-power plasma cleaning.

### Reflectance measurements

The substrates were placed in a custom-designed sample holder, sealed with a FastWell silicone gasket and covered with a clear borosilicate glass slide. The nanoarrays were immersed in PBS by injecting the buffer into the cavity of the holder. The sample was mounted on the stage of a custom-built Stokes polarimeter. Incident polarisation was adjusted using a linear polariser, and optical rotation dispersion (ORD) was measured by capturing intensities at four analyser angles: 0°, 45°, 90°, and 135°. Reflectance data was recorded with the analyser set at 0°.

### Mueller Matrix Polarimetry

For Mueller Matrix Polarimetry (MMP) measurements, samples were prepared in the same manner as for Stokes polarimetry. The highly collimated beamlight of B23 beamline at Diamond Light Source enabled the use of the MMP that consist of two pairs of photoelastic modulators flanked the sample, serving as the polarisation state generator and analyser[37].This setup enables the production and measurement of all required light states to determine the differential 16 Mueller matrix elements. The LD, LD’, LB, LB’, CD, and CB were calculated using the Analitic Inversion method [38].

**Numerical Simulations**

Simulations were conducted using the commercial finite element analysis software COMSOL Multiphysics v6.1, specifically the Wave Optics module. The nanostructure was modelled within a cuboid unit cell, where the x and y dimensions defined the metamaterial’s periodicity as determined from AFM images. The z dimensions of the cell were sufficiently large (≥λ max/2) to ensure that near-fields from the nanostructures did not interact with integration surfaces above the structure; the total height of the cell was set to 1800 nm. The unit cell was divided into domains of different thicknesses, with the top and bottom 200 nm domains serving as perfectly matched layers (PMLs) to absorb all reflections. The boundary adjacent to the upper PML acted as the excitation port, from which the incident light was directed and its polarization specified. Reflection intensity was also calculated at this port. The silicon structure was situated near the centre of the cuboid, while the boundary adjacent to the lower PML functioned as the outgoing port.

# Conflicts of Interest

There are no conflicts to declare.

# Author Contributions

The manuscript was written through contributions of all authors. All authors have given approval to the final version of the manuscript.

# Acknowledgements

The authors acknowledge financial support from the Engineering and Physical Sciences Research Council (EP/S012745/1 and EP/S029168/1). Technical support from the James Watt Nanofabrication Centre (JWNC) and Diamond Light Source Ltd. is acknowledged. DKM was awarded a studentship by the EPSRC. M.K. acknowledges the Leverhulme Trust for the award of a Research Fellowship (RF-2019-023).

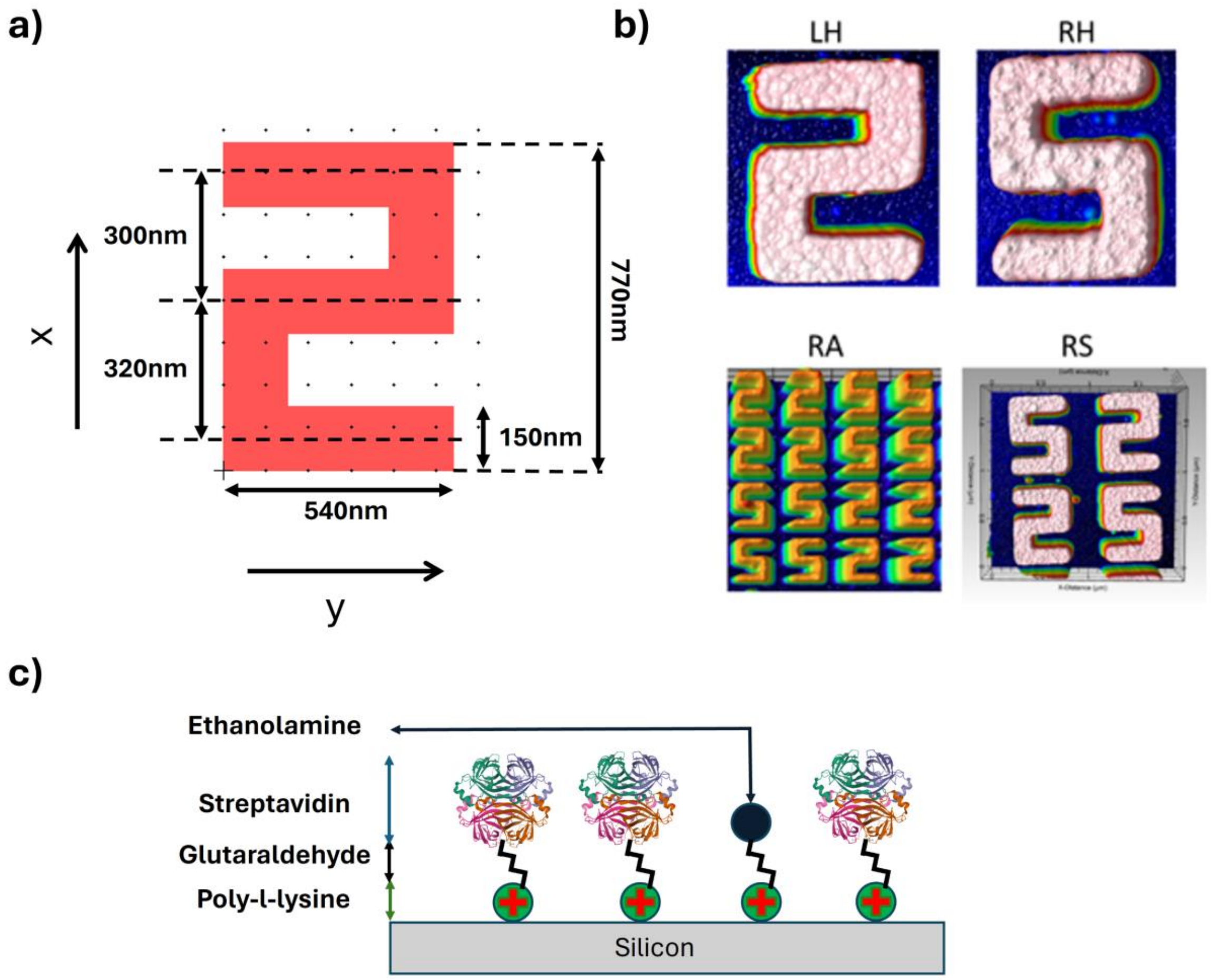

**Figure 1.** a) A schematic of a single LH S-structure with labelled dimensions. Black dots represent 0.1 µm intervals. b) AFM images from each array type, with the RA image taken from the intersection of four 19x19 arrays. c) A schematic of the deposition process with labelled constituent layers. Positively charged poly-L-lysine binds to the silicon surface. Glutaraldehyde crosslinks the lysine to streptavidin. Ethanolamine serves as a capping layer to limit nonspecific binding, allowing streptavidin to bind with additional species.

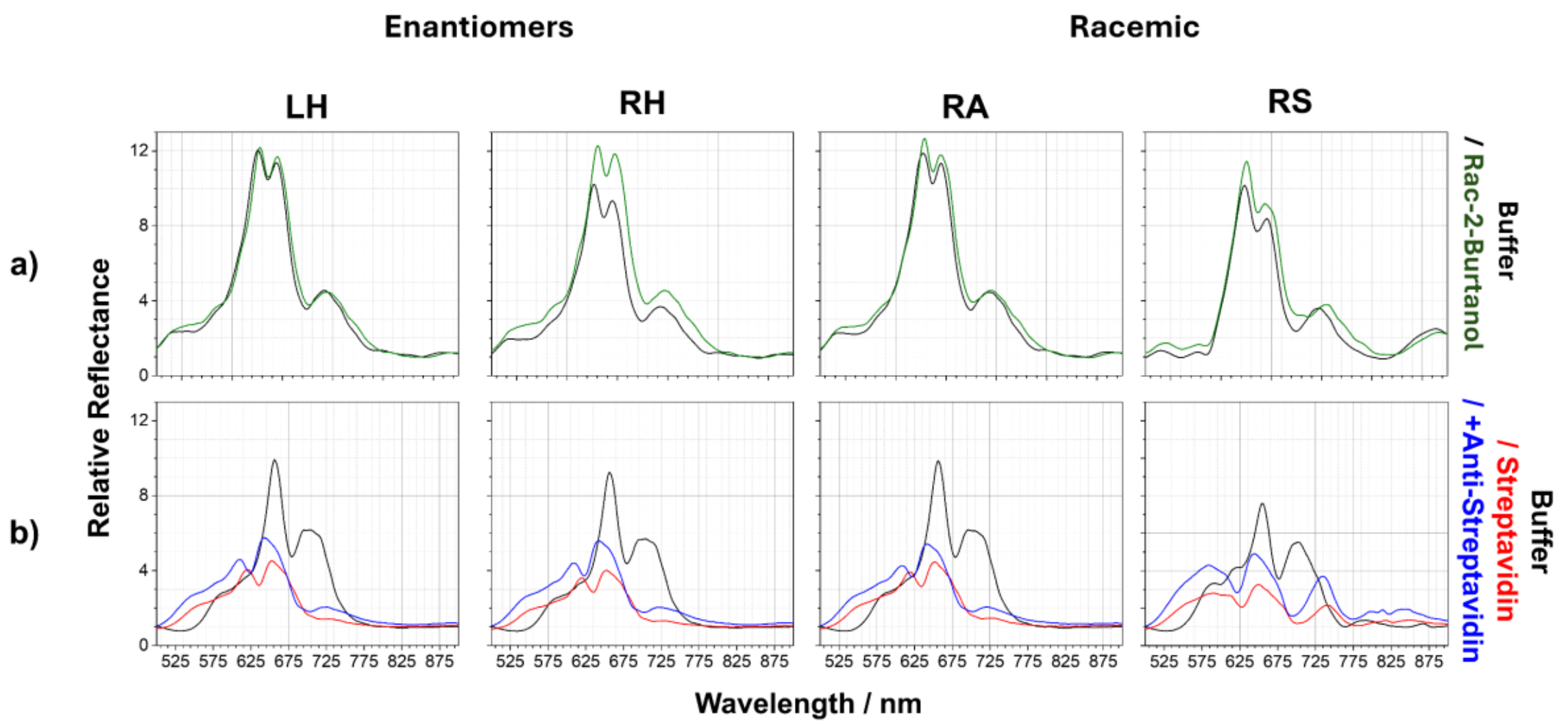

**Figure 2.** a) Reflectance spectra from a structure with a nominal thickness of approximately 160 nm, showing the effects of changes in refractive index when transitioning from water to rac-2-butanol. b) Experimental y-polarized reflectance spectra collected in PBS buffer for LH, RH, RA, and RS arrays of a ~160 nm sample (different from the one used in a). The spectra are shown for the buffer (black), after the deposition of streptavidin (red), and after the deposition of anti-streptavidin polyclonal antibody (blue).

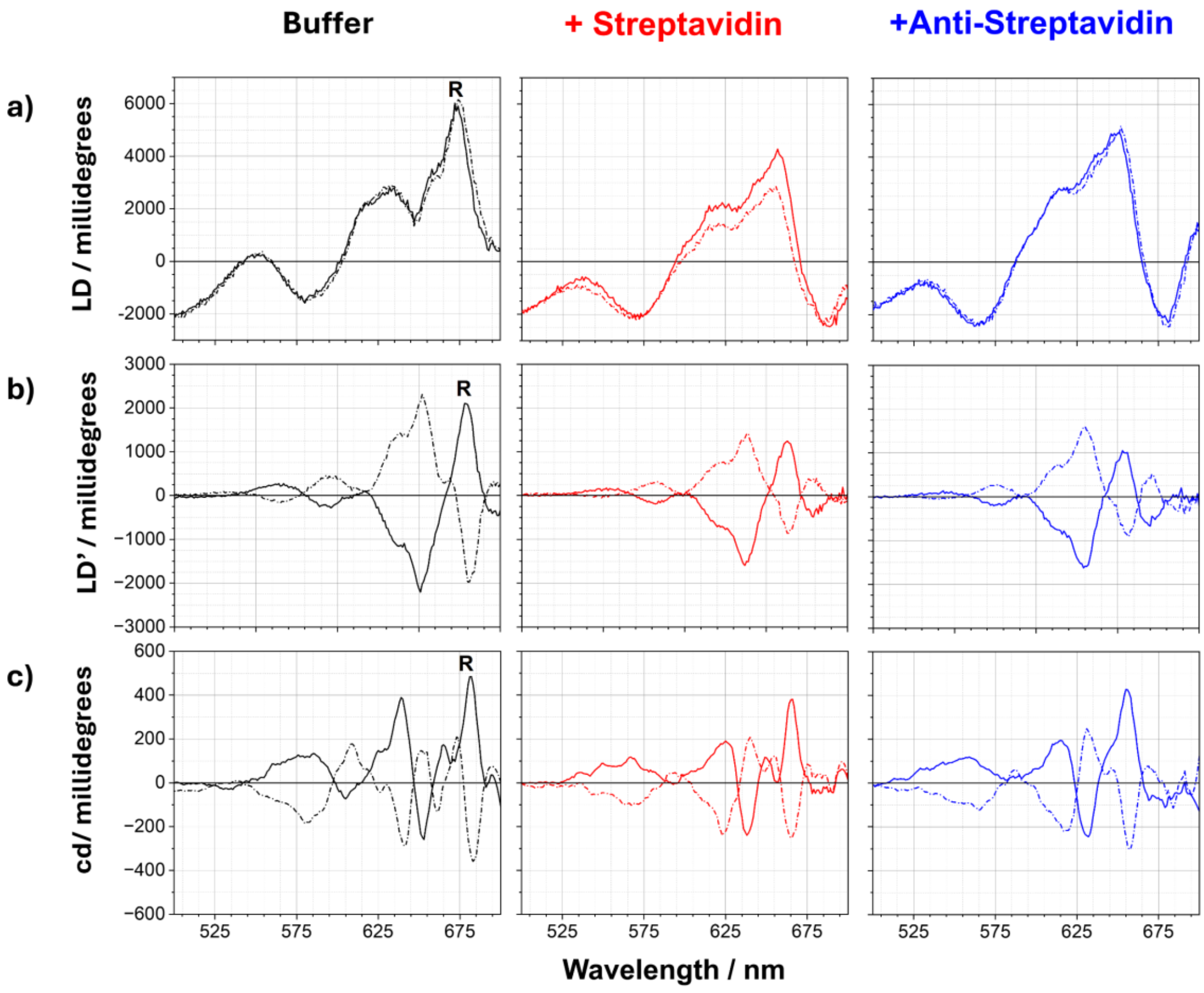

**Figure 3.** a) MMP LD spectra for a sample with a nominal thickness of approximately 210 nm, showing LH (solid) and RH (dashed) structures in PBS buffer: no biomolecule (black), with streptavidin (red), and with anti-streptavidin (blue) deposited. b) MMP LD' spectra for the same sample with the same conditions. c) MMP CD spectra for the same sample with the same conditions. The asymmetry parameter displayed in table 1 are derived from the peak labelled R. The MMP LD, LD' and CD spectra were calculated using the Analytic Inverse method (reference of Arteaga et al).

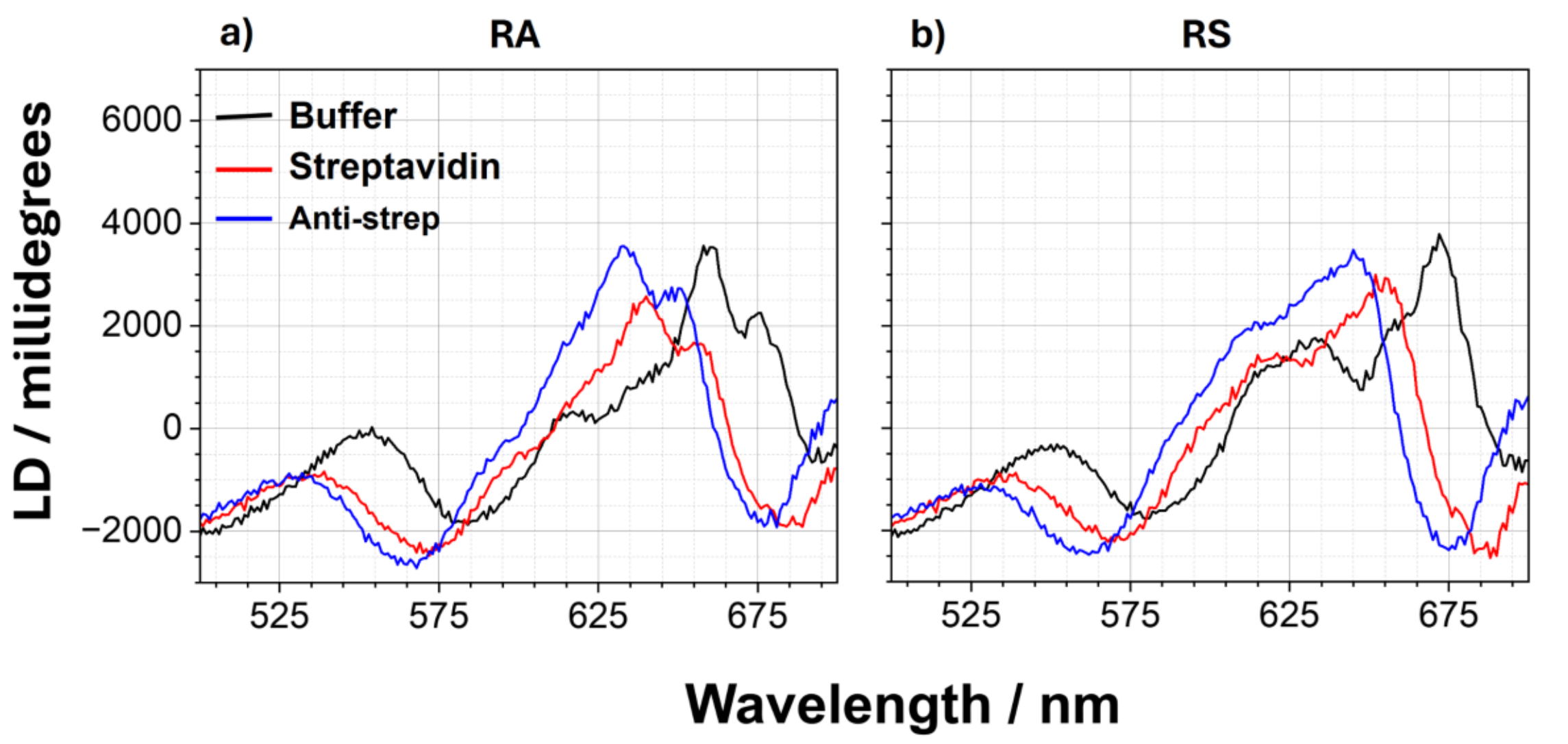

**Figure 4**. LD spectra for a) RA and b) RS samples with nominal thicknesses of approximately 210 nm, showing LH (solid) and RH (dashed) structures in PBS buffer: no biomolecule (black), with streptavidin (red), and with anti-streptavidin (blue) deposited. The asymmetry parameter displayed in table 2 are derived from the peak labelled R.

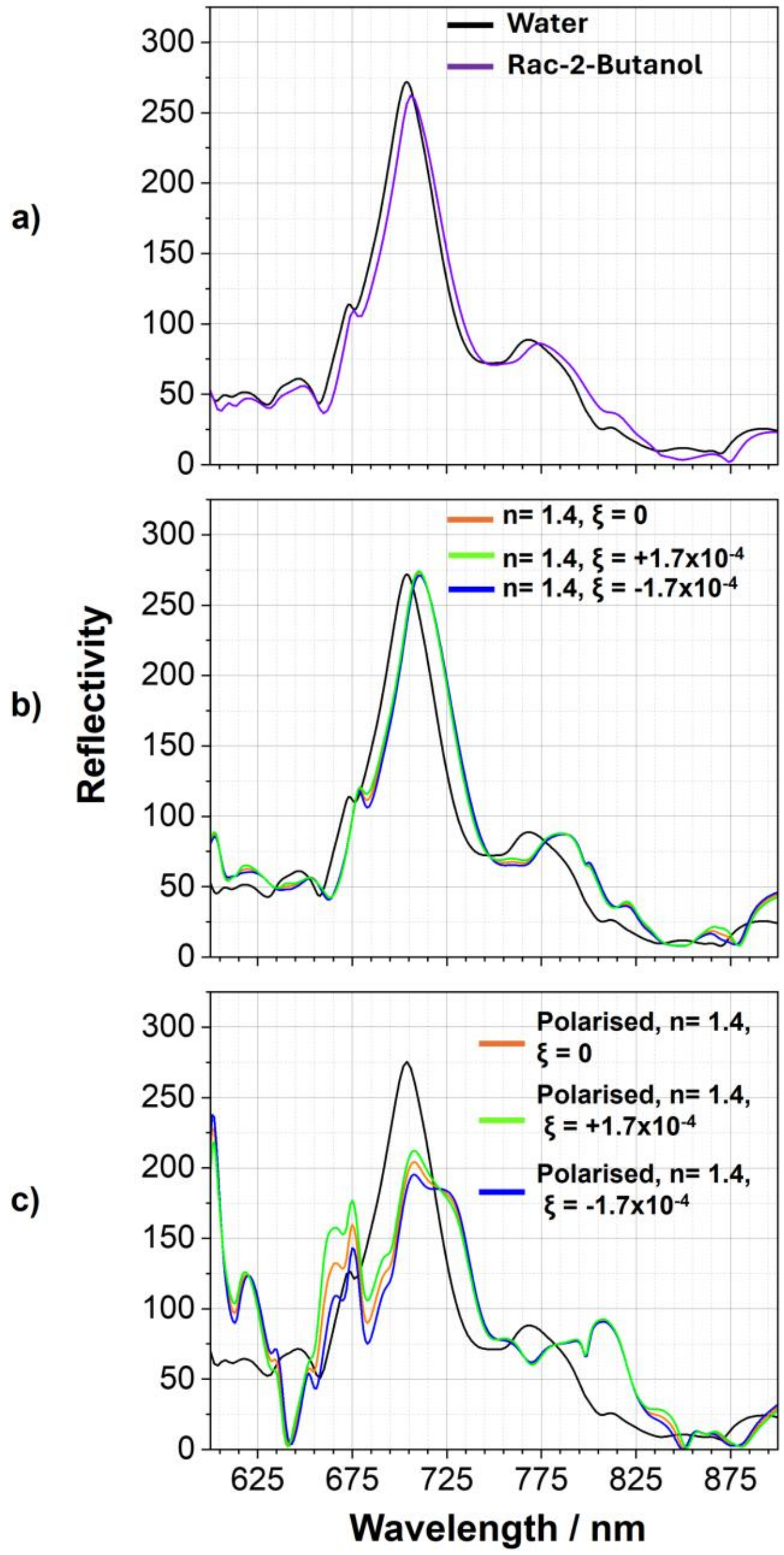

Figure 5. Numerical simulations for: a) an LH structure immersed in water and rac-2-butanol; b) an LH structure immersed in buffer with a chiral dielectric layer; and c) an LH structure immersed in buffer with a polarized chiral dielectric layer. The simulations with a negative Pasteur coefficient (ξ) are symmetry equivalent to an RH structure with a positive ξ.

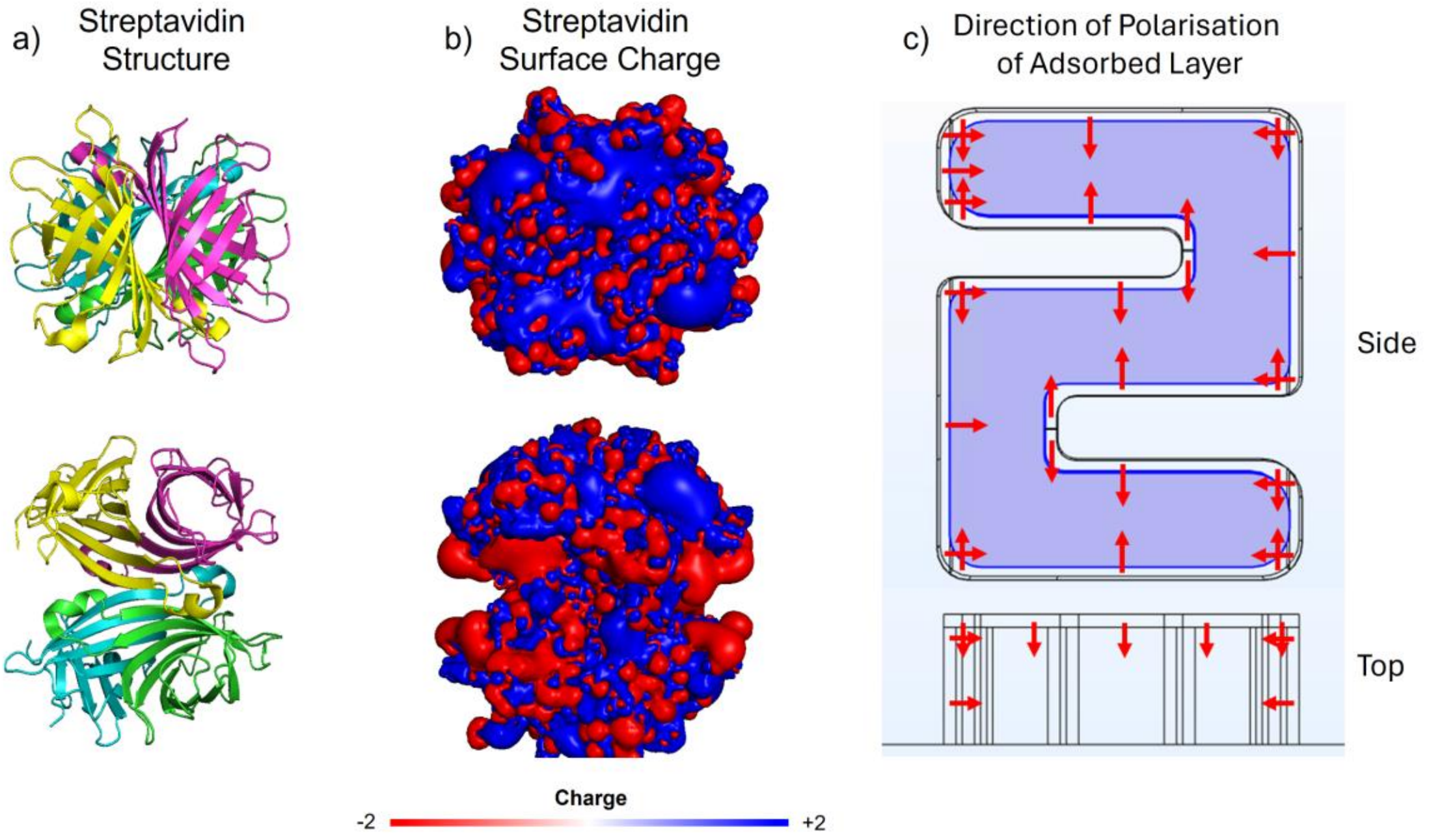

**Figure 6** a) A ribbon representation and b) PyMOL generated electrostatic potential map of streptavidin from two perspectives (see S4 Supplementary Information). c) Model used in numerical simulations, displayed from side and top views, indicating the direction of the dipole layers.

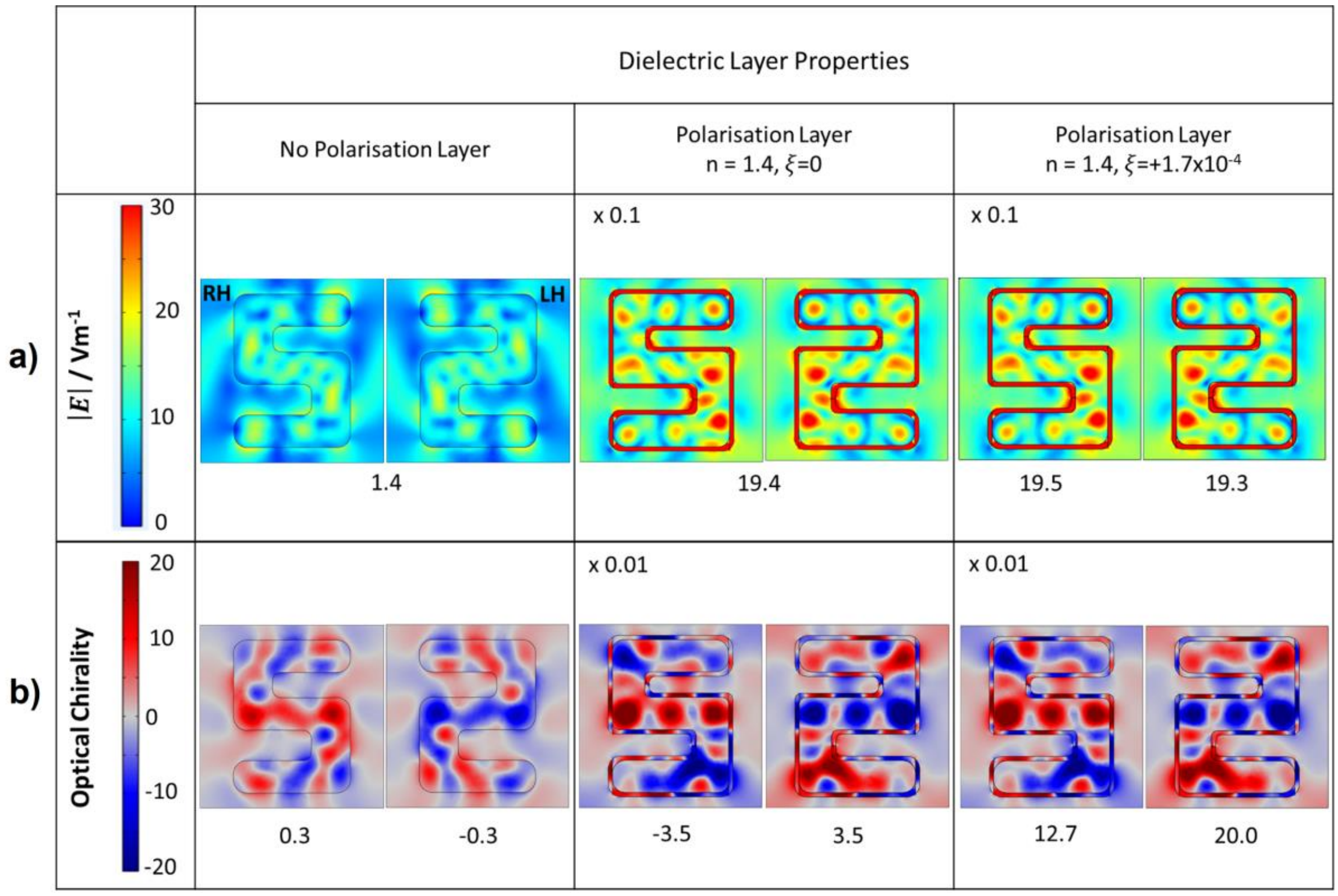

**Figure 7.** Comparison of a) |E| and b) optical chirality maps derived from numerical simulations. The left column shows the structure without an external layer, the middle column with an achiral polarized layer, and the right column with a chiral polarized layer. Surface-integrated values for the maps (taken from the midpoints of the structure) are labelled beneath the respective images. The scales for the polarized layers are adjusted and labelled in the top left corner of the middle column.

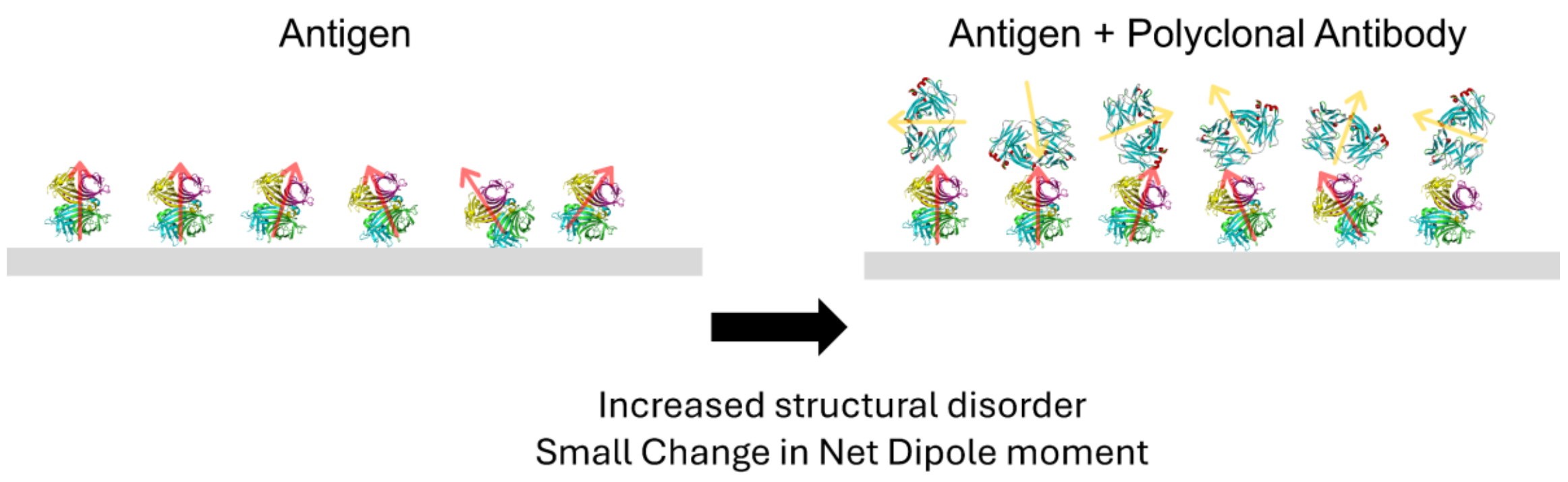

**Figure 8.** An illustration showing the relative levels of structural order for streptavidin (antigen) and streptavidin - anti-strept complex.

| | Asymmetry Parameter | Values |
|---|---|---|
| CD | $\alpha_{Buffer}^{CD}$ | $1.37 \pm 0.04$ |
| | $\alpha_{Strept}^{CD}$ | $1.56 \pm 0.04$ |
| | $\alpha_{Anti}^{CD}$ | $1.42 \pm 0.04$ |
| LD' | $\alpha_{Buffer}^{LD'}$ | $1.07 \pm 0.02$ |
| | $\alpha_{Strept}^{LD'}$ | $1.44 \pm 0.02$ |
| | $\alpha_{Anti}^{LD'}$ | $1.18 \pm 0.02$ |
| LD | $\alpha_{Buffer}^{LD}$ | $0.99 \pm 0.02$ |
| | $\alpha_{Strept}^{LD}$ | $1.51 \pm 0.02$ |
| | $\alpha_{Anti}^{LD}$ | $0.97 \pm 0.02$ |

**Table 1.** Asymmetry parameters derived from LD, LD', and CD spectra for unfunctionalized, streptavidin, and anti-LH and RH substrates, as shown in Figure 3.